\documentstyle[epsfig]{aipproc}

\begin{document}
\title{Derivation of a Sample of\\
Gamma-Ray Bursts from\\
BATSE DISCLA Data}

\author{Maarten Schmidt}
\address{California Institute of Technology, Pasadena, California 91125}

\maketitle

\begin{abstract}
We have searched for gamma-ray bursts (GRBs) in the BATSE DISCLA data
over a time period of 5.9 years. We employ a trigger requiring an 
excess of at least $5\sigma$ over background for at least two modules
in the $50-300$ keV range. After excluding certain geographic locations
of the satellite, we are left with 4485 triggers. Based on sky positions,
we exclude triggers close to the sun, to Cyg X-1, to Nova Persei 1992
and the repeater SGR $1806-20$, while these sources were active. We
accept 1013 triggers that correspond to GRBs in the BATSE catalog,
and after visual inspection of the time profiles classify 378 triggers
as cosmic GRBs. We denote the 1391 GRBs so selected as the ``BD2 sample''.
The BD2 sample effectively represents 2.003 years of full sky coverage
for a rate of 694 GRBs per year. Euclidean $V/V_{max}$ values have been
derived through simulations in which each GRB is removed in distance
until the detection algorithm does not produce a trigger. The BD2 sample
produces a mean value $<V/V_{max}> = 0.334 \pm 0.008$.

\end{abstract}

\begin{figure}[t!]
\centerline{\epsfig{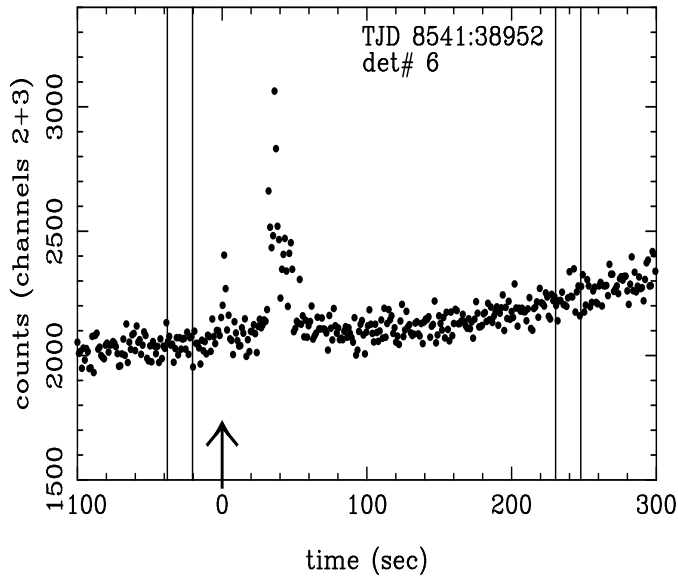}}
\vspace{10pt}
\caption{The test for the presence of a burst in time interval 
$0.000-1.024$ sec involves a linear interpolation between two 17.408 sec
averages of the background, one ending at time $-20.48$ sec, the other
starting at time $230.4$ sec.}
\label{F:GP-37:1}
\end{figure}

\section*{Introduction}

We have been engaged for several years in an effort to derive a  
homogeneous sample of gamma-ray bursts (GRBs) from the continuous data
stream transmitted by the Burst and Transient Source Experiment (BATSE).
In particular, we have used the DISCLA data which provide the counts
for each of the eight BATSE detectors in channels $1-4$ on a time scale
of 1024 msec. The resulting {\it BD2 sample} of 1391 GRBs covers a time 
period of 5.9 years \cite{schmidt99a,schmidt99b}. It includes 378 GRBs 
that are not in the BATSE catalog maintained on the BATSE web site.
The BD2 sample has been used for a derivation of the characteristic 
luminosity and local space density of GRBs \cite{schmidt99b}. 

We will briefly review the procedures used to construct the BD2 sample,
and analyze the differences between the BATSE catalog and the BD2 sample
in some detail.

\section*{Derivation of the BD2 Sample}

For most of the time period covered by the BATSE catalog,
the on-board trigger mechanism required that the counts in
channels 2 and 3 covering the energy range $50-300$ keV exceed the
background by 5.5$\sigma$ on a time scale of 64, 256 or 1024 msec in 
two of the eight detectors. In our search based on DISCLA data, we
used channels 2$+$3, and required an excess of 5$\sigma$ above background
on a time scale of 1024 msec in two detectors.

The BATSE on-board trigger employs a background averaged over 17.408 sec
that is updated every 17.408 sec. We have taken advantage of the 
archival nature of the DISCLA data to use a background that is derived
by linear interpolation from two averages taken over 17.408 sec, as
shown in Figure \ref{F:GP-37:1}. The interval of 20.48 sec between
the first background average and the test time is intended to alleviate
the problem of detecting slowly rising GRBs \cite{hig96}.

\begin{figure}[t!]
\centerline{\epsfig{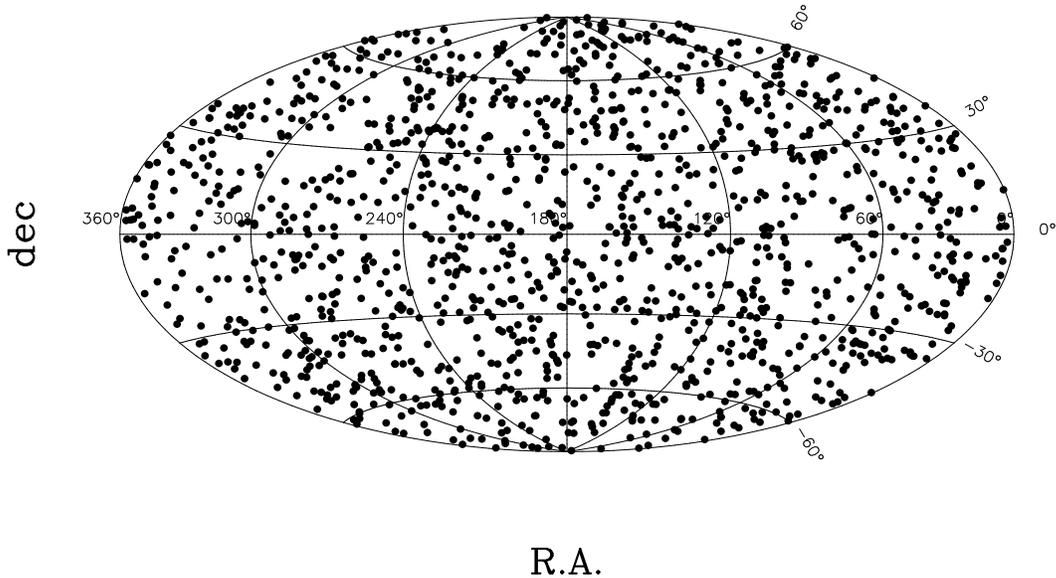}}
\vspace{10pt}
\caption{Equatorial coordinates of 1391 GRB in the BD2 sample}
\label{F:GP-37:2}
\end{figure}

We searched for defects in the DISCLA data, since these could lead to
false triggers or affect the background. For the time period of
TJD 8365-10528, we found around 151,000 defects, ranging from checksum
errors that affected only one 1024 msec bin to gaps caused by
transmission problems, passage through the South Atlantic Anomaly, etc.
Around each of these defects, we set up an exclusion time window such
that the defect does not affect the background.

The initial search yielded 7536 triggers. The geographic coordinates
of the satellites at the time of trigger showed strong concentrations
over W. Australia and over Mexico and Texas \cite{schmidt99a}. We then
outlined geographic exclusion zones to avoid the trigger concentrations.
With these exclusions in place, we were left with 4485 triggers. For 
each of these we derived celestial coordinates based on the relative
response in all eight detectors and the orientation of the satellite,
and other properties such as the duration, the hardness ratios, 
$V/V_{max}$, etc.

The equatorial coordinates of these triggers showed clear concentrations
that were identified as Cyg X-1, Nova Persei 1992, and solar flares
along the ecliptic \cite{schmidt99a}. We excluded from consideration
all triggers whose positions were within 23 deg of these sources while
they were active. Most of the remaining triggers were either
magnetospheric events or cosmic GRBs. For a more detailed description
of the selection procedure, the reader is referred to \cite{schmidt99a}. 
We ended up with a sample of 1422 GRBs \cite{schmidt99a} which we call 
the BD1 sample.

\begin{figure}[t!]
\centerline{\epsfig{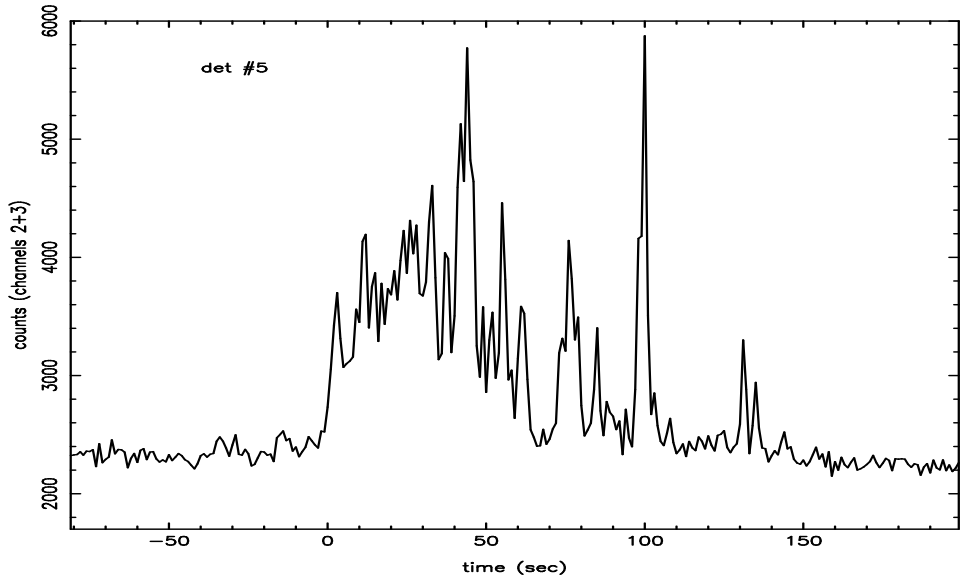}}
\vspace{10pt}
\caption{Example of a gamma-ray burst in the BD2 sample
(TJD 8868:05731) that is not listed in BATSE catalog}
\label{F:GP-37:3}
\end{figure}

Subsequently, we investigated carefully all our bursts that were either 
not in the BATSE catalog (for an example, see Fig. \ref{F:GP-37:3}), 
or whose positions or times agreed poorly with the catalog data.
We eventually rejected 31 of the sources in the BD1 sample (most of which
were parts of long bursts or identified as the repeater SGR $1806-20$), 
resulting in the BD2 sample of 1391 GRBs \cite{schmidt99b}. We show in 
Table 1 an updated accounting of the classification of all 4485 triggers.

The most important property among those derived for the GRBs in the BD2 sample 
is $V/V_{max}$. Since redshifts are not know for most GRBs, $V/V_{max}$
has to be evaluated in euclidean space. It has usually been assumed that
$V/V_{max} = (C_{max}/C_{min})^{-3/2}$, where $C_{min}$ is the minimum
detectable burst signal, and $C_{max}$ the maximum amplitude.
Instead, we derive $V/V_{max}$ of each GRB by carrying out a simulation 
in which we move the source
out in euclidean space, and at each step apply the detection algorithm
to see whether the reduced burst is still detected
\cite{schmidt99a}. In the process
of moving out the source, it may get detected later and later (depending on
its time profile) and some burst signal may be included in the background.
In some cases, the $C_{max}$ part of the profile is never detected
before the source disappears as it is being removed. Using $C_{max}$ to
derive $V/V_{max}$ therefore tends to lead to an underestimate of
$V/V_{max}$.

Using the procedure outlined above, the BD2 sample produces a mean value 
$<V/V_{max}> = 0.334 \pm 0.008$ \cite{schmidt99b}. The deviation from 
the value 0.5 expected for a uniform space distribution reflects to
first order the effect of using euclidean space in its derivation rather
than a relativistic cosmological model. Hence the euclidean $<V/V_{max}>$
is effectively a distance indicator, allowing derivation of the
characteristic luminosity of GRBs\cite{schmidt99b}.

\begin{table}
\caption{Classification of 4485 triggers}
\label{table1}
\begin{tabular}{lllllrr}
   Description&&&&& reject & accept\\
\tableline
Within 23 deg of the sun when active     &&&&& 963 &      \\
Within 23 deg of CygX-1 when active      &&&&& 827 &      \\
Within 23 deg of Nova Persei when active &&&&& 418 &      \\
Within 230.4 sec of BATSE-listed burst   &&&&&     & 1013 \\
Profile inspected: accepted as GRB       &&&&&     &  378 \\
Profile inspected: rejected              &&&&& 818 &      \\
Part of a preceding long burst           &&&&&  18 &      \\
Soft repeater $1806-20$                  &&&&&   6 &      \\
Near sun, soft spectrum                  &&&&&  44 &      \\
\tableline
Total number of GRBs in the BD2 sample:  &&&&&     & 1391 \\
\end{tabular}
\end{table}

\section*{Comparison of the BATSE catalog and the BD2 sample}

Given that the BD2 sample was derived independently of the 
BATSE catalog, it is of interest to compare the two data sets.

1) The BD2 sample produces an all-sky rate of 694 per year, the BATSE
catalog yields $\sim$ 690 per year \cite{meegan98}. One might expect a 
larger rate in the BD2 sample since its limiting S/N is 5.0, while for 
the BATSE catalog it is 5.5. However, the BD2 sample is limited to
detections at the time scale 1024 msec, while the BATSE catalog also 
includes time scales of 256 and 64 msec.

2) The BD2 sample has 378 GRB not in the BATSE catalog, and the BATSE 
catalog has 130 GRB, detected at a time scale of 1024 msec, that are 
not in the BD2 sample. These differences are related to the 
different depth of the two data sets (5$\sigma$ vs. 5.5$\sigma$), 
but they are also influenced by items (3)--(6) below.

3) The $\sim$ 151,000 time exclusion windows used in the derivation of
the BD2 sample are independent from those used in the BATSE search.

4) Following a BATSE trigger, the detection mechanism was disabled
until data could be transmitted to the ground. After a BD2 detection,
we disabled the software trigger for 230 sec.

5) The trigger criteria for the BATSE database were changed a number
of times, see \cite{meegan98}.

6) In general, if the background stretches used in two searches are 
different, then if a source is detected precisely at the limiting S/N 
in one of the searches, the probability that the other search finds the
source is around 50\%. The backgrounds used in the BATSE search and
the BD2 sample are independent when the BATSE background stretch is
separated by less than 3 sec from the time of detection and partly
dependent when the separation is larger than 3 sec. Since we would
find a substantial number of GRBs that are not in the BD2 by 
carrying out a search using different background stretches than 
those of Figure \ref{F:GP-37:1},
we prefer to call the resulting collection of sources
a sample. It should be stressed that
the different samples so found are all statistically equivalent to each
other, as long as one background choice is not 
to be preferred over the other.

\end{document}